\documentclass[pre,aps,preprint,amssymb]{revtex4}

\pdfoutput=1

\usepackage{amsmath}
\usepackage{tabularx}
\usepackage{pifont}
\usepackage{makeidx}
\usepackage{epsfig}
\usepackage{dcolumn}

\begin{document}

\title{Diffusion across a concentration step: Strongly nonmonotonic evolution
into thermodynamic equilibrium}

\date{\today}
\author{Hans R. Moser \\
{\small Physik-Institut, University of Z\"urich,}
{\small Winterthurerstrasse 190, CH-8057 Z\"urich, Switzerland} \\
{\small E-mail: moser@physik.uzh.ch}}

\begin{abstract}
Dynamical and statistical behavior of the ionic particles in dissolved salts
have long been known, but their hydration shells still raise unsettled questions.
We engineered a ``diffusion tunnel diode" that is structurally analogous to the
well-known Esaki diode, but now concentration gradients serve as generalized
voltages and the current means particle flow. In an equipartition sense, the
hydrated ions enter a cavity as individual particles and later, upon increase of
their concentration therein, they lose water molecules that henceforth are
particles of their own. These temporarily attached water molecules thus are the
tunnel current analogue. Unlike the original tunnel diode, our negative
differential resistance has implications for the second law of thermodynamics,
due to thermal effects of changes in the hydration shells.

\vspace{5mm}
\noindent
Keywords: Generalization of tunnel diode; Negative differential resistance;
Properties of hydration shells; Equipartition law; Entropy

% \vspace{5mm}
% \noindent
% PACS numbers:

\end{abstract}

\maketitle

\section{Introduction}
The transfer of electrical conduction mechanisms to situations with other driving
forces has a long-standing tradition. Heat conduction driven by temperature
gradients, downhill motion of water due to gravity, and diffusion caused by
concentration gradiens (our present instance) are common examples. A somewhat
less intuitive case is magnetism, where field strength, flux density, and
permeability correspond to driving force, current density, and conductivity,
respectively.

Usually such issues are studied in a linear regime in the spirit of a generalized
Ohm's law. However, the tunnel diode \cite{esaki} departs dramatically from
linearity, due to a current-voltage characteristic that involves
a range of negative differential resistance $dU/dI$ that is able to
undamp an oscillatory circuit. Clearly, this cannot end up in
a resonance catastrophe, since amplification stops as this range
gets exceeded by the current or voltage amplitude. In any case,
there is also a positive static (that is, ohmic) resistance contribution that
dissipates energy from the power supply.

As indicated previously, we intend to translate the essentials of the tunnel diode
to the case where potential levels are established by particle concentrations,
giving rise to diffusive particle currents. At first glance this is
doomed to failure, since the tunnel current has no classical analogue
(and we disregard de Broglie wave tunneling of particles as heavy as hydrated
ions). We pursue quite another course by looking for a purely structural
correspondence that in substance is entirely different. Sometimes such analogues
are said to be isomorphic.

In the description of the experiment we shall see how the additional particle
flow (beyond ordinary diffusion) comes about. We choose NaCl for practical
reasons, among them its high solubility in water. Essentially, we consider ions
that complete their hydration shell outside of a cavity and later they release the
water molecules inside again. In short, we use that there is maximum degree
of hydration at minimum concentration of the ions. The occurrence of a negative
differential resistance then is straightforward, since the ``voltage" between
a reservoir with saturated solution and the interior of said cavity is at
its maximum for zero ionic particle concentration therein (see below that we
additionally organize a zone of low salt concentration, thus enhanced hydration,
right outside the cavity). But then there is not yet any water molecules
release, and so the transfer rate of separate (independent) water molecules
into the cavity peaks at some intermediate voltage. We are aware of other
contributions that involve negative differential resistances as well, sometimes
also in quite a generalized context with respect to the original Esaki diode,
we mention the experiments \cite{renk,kannan}. Further, Ref. \cite{li} presents
a simulation not too far from a possible experimental implementation, where
the driving force is no longer an electric field strength but a temperature
gradient. However, all these authors largely focused on issues we do not primarily
have in mind.

The delicate behavior of hydration shells is already a worthy research subject
in our view, but we have an even more ambitious goal. Due to energy conservation,
hydration is attended with thermal effects, but this concerns the (generalized)
tunnel current only. Whereas, in terms of the equipartition law, the ``ordinary"
particle current, that is, the flow of ionic carriers for the polar water
molecules is just made of individual particles, regardless of their hydration
status. This has exciting consequences that we intend to investigate.

\section{Setup for generalized tunnel diode}
A device with the desired properties must necessarily have an initial particle
concentration gradient, a barrier of geometrically restricted (or, conversely,
admitted) moderate diffusion conductance in order to separate regions of different
concentrations, and also a possibility to restore the initial situation. Then,
very importantly, there should be a built-in equipment to measure small
temperature differences and thus, by virtue of the hydration heat, indirectly also
concentration profiles.

Fig.~1 presents the required items and, we think, their meaningful assembly,
as will emerge later. Bottom-up we see a petri dish that serves as the
reservoir for saturated NaCl solution, a toothed structure that stands for
a grid to establish a clearance to the thermoelectric generator (TEG) just above,
and then a porous glass fiber paper for the diffusion barrier already mentioned.
Then there is a plastic frame with a square hole, covered with PTFE tape
(Polytetrafluoroethylene, often referred to as Teflon). Above that we use
a glass plate that carries a funnel containing glass granules soaked with water
by capillary forces, and also the hole or cavity underneath is filled with plain
water. All of this is fairly well isolated from the environment by simple means,
as only temperatures very close to ambient occur (see below), and so there
is little heat exchange. Uppermost there is another funnel for (optional) water
inlet, i.e., to flush or replace the water in the aforesaid square cavity, as
will be explained later. But this new external water just enters the lower funnel,
while flushing takes place with temperature-adapted water that has been there for
a long time. Moreover, the water marginally tends to leak out between plastic
frame and the moderately hydrophilic glass plate, but the Teflon layer there
reliably prevents that.

\begin{figure}

\vspace{-8.0cm}

\hspace{-1.0cm}\includegraphics[scale=0.83, angle=0]{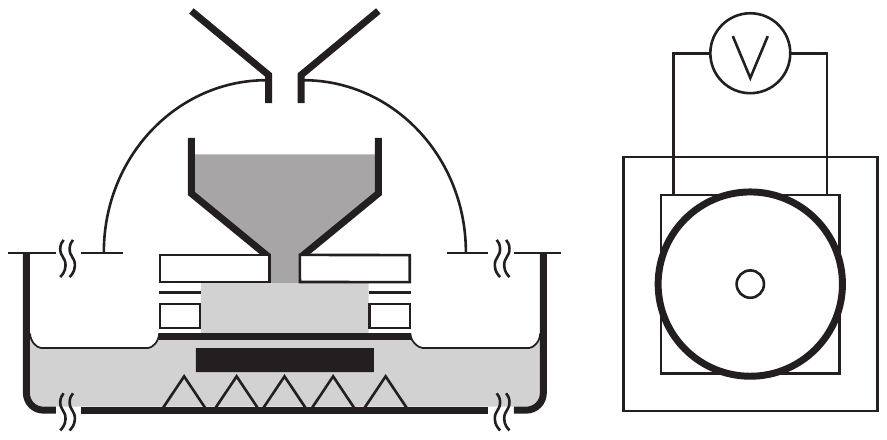}

\vspace{-13.2cm}

\caption{Left: Quasi-realistic illustration of the entire setup, see text. Not all
of the proportions are completely realistic, as we optimize for intelligibility
rather than practical construction details. In particular, we slightly tend towards
an ``exploded view" to recognize the stacked items. Right: Top view restricted to
selected parts, namely glass plate that supports the lower funnel, top rim
of either funnel, as well as the thermoelectric generator (TEG) connected to
a voltmeter.
\label{fig1}}

\vspace{0.3cm}

\end{figure}

The commercial TEG is a (waterproof sealed) 13 by 13~mm square with an efficiency
of some 0.06~V/K, and the hole in the plastic frame is around the same area
and 1.5~mm in height. Such lengths matter primarily because of
diffusion velocities, as we shall see in the measurements section. This also
means that only within certain limits we might proportionally enlarge
our setup. Further, the glass fiber paper provides the barrier or obstacle
to diffusion from the NaCl solution reservoir into the cavity
inside the plastic frame. We use a filter paper around 0.45~mm thick with 93\%
porosity and 1.5~$\mu$m pore size in terms of particle retention, but this
turns out to be surprisingly uncritical. To see that, we tried greatly different
substitutes, among them metal meshes of various thickness and porosity,
which altered things just quantitatively. In this respect at least, our setup
is quite robust.

With these preliminaries we may now anticipate what our particular arrangement
in Fig.~1 is good for, why it is the analogue of a tunnel diode and what we
actually want to measure. We place the actual device (with lower funnel and
cavity in the plastic frame filled with plain water) into the prepared saltwater
reservoir, assemble the isolation from the surroundings and connect the TEG
to the voltmeter. Then, at the very beginning of what follows, the temperatures
at the TEG should be as balanced as possible (that is, little offset in any sign).
Once all the material is provided, this is most easily accomplished  by trial and
error, as recording such TEG voltage offsets is not overly elaborate. We
have already briefly outlined what then happens, namely diffusion out of the
reservoir into the cavity, attended with thermally relevant hydration effects.
Outside the cavity, predominantly in the filter paper out there (that has also
been previously soaked with plain water), we get a ``blurred zone" with
by far less salt concentration than in the reservoir. There, the hydration
increases significantly with respect to the reservoir, which causes heat
(or enthalpy) of dilution. Such dilution processes occasionally can even
be violent, as is known from the dilution of acids. The reverse then happens
in the cavity; the concentration therein grows over time, causing the
hydration to decrease and hence the temperature to drop. Just this mechanism will
ultimately be used in the subsequent figures. In our view, it is instructive
to recognize that in saturated NaCl solution there are only some 4.5 times more
water molecules than ions, and so these ``few" molecules must account for both
hydration as well as liquid between the ions. Provisionally we only note that
most likely every water molecule gets involved in hydration, at least temporarily.

Further, the geometry in Fig.~1 deserves some remarks in order to understand the
tunnel diode analogy. The hydrated particles diffuse into said cavity from the
outside and then proceed further inwards. Clearly, the generalized potential
difference is largest between the reservoir and the innermost part of the cavity.
Here the intended analogy is certainly less intuitive than in a geometry where
the particles perform a transit, i.e., enter the cavity on one side and exit on
the opposite one. However, this causes significant engineering problems, and so
we stick to the more stable and much simpler version of Fig.~1. We merely have
to accept a minor additional generalization step with regard to the archetypal
electronic tunnel diode.

\section{Measurements and their physical meaning}
First, we highlight some key properties of diffusion, since they are supremely
relevant to experimental geometries in the spirit of Fig.~1 due to rates of
change in particle concentrations. At the same time, we also want to look
at the Brownian motion of individual particles, primarily because together they
make up diffusion as a whole, and also regarding the hydration effects that rather
incompletely may be described as a macroscopic overall phenomenon. The interplay
of many particles in statistical mechanics and thermodynamics is a vast
research domain, we refer to \cite{kubo,blundell} for many relevant aspects. We
start out with the Langevin equation for a single particle
\begin{equation}
m\ddot{{\bf x}}(t) = -\beta\dot{{\bf x}}(t)+{\bf F}_c(t)+{\bf F}_{ext}
\end{equation}
that, besides friction term and possible external force, comprises the random
driving force ${\bf F}_c(t)$ due to collisions. Frequently, external
force means gravitation, but here this is negligible since the hydrated
Na$^+$ and Cl$^-$ ions do not sediment to the ground. Also, we do not apply
an electric field. The collision force is thought to be a long-term function of
time, i.e., an irregular sequence of many different collisions.

This situation may be reconsidered as a momentary view of the sum over many
particles that perform a random walk according to Eq.~(1), once we accept
ergodicity to apply in this case. Conveniently, isotropy removes the random term
from this sum, and as a final prerequisite we make use of equipartition that
introduces temperature. This way, after some steps of rearrangement we arrive at
our preferred version of Eq.~(1), namely an equation that also covers the few just
above stated clear-cut assumptions. This reads
\begin{equation}
\frac{d^2}{dt^2} \langle {\bf x}(t)^2 \rangle_x
+ \frac{\beta}{m} \frac{d}{dt} \langle {\bf x}(t)^2 \rangle_x
= \frac{2k_B T}{m},
\end{equation}
where the spatial mean value of squared distances the particles have departed
(in some fixed direction) from their starting point is our new variable. The
stationary solution of Eq.~(2) is a fairly known result, namely the linearly
growing mean value
\begin{equation}
\langle {\bf x}(t)^2 \rangle_x = 2Dt, \qquad D = \frac{k_B T}{\beta}.
\end{equation}
Here $D$ is the diffusion constant as it enters the two major statements for
unconstrained diffusion (in large volume liquids) according to Fick's laws, namely
\begin{equation}
{\bf j}_n({\bf x},t) = -D\nabla n({\bf x},t), \qquad
\frac{\partial n({\bf x},t)}{\partial t} = D\Delta n({\bf x},t).
\end{equation}
The vector ${\bf j}_n({\bf x},t)$ denotes the particle current density, and
Eqs.~(4) describe Brownian motion as an overall diffusion phenomenon of many
particles with density $n({\bf x},t)$. We recognize that already unrestricted
diffusion constitutes a quasi-equilibrium of Brownian particles with their host
medium, since there exists a macroscopic time-evolution while the density
gradients tend to even out. Admittedly the process is slow, and so the
equipartition law stays almost unaffected. Apart from the time-evolution of
$n({\bf x},t)$ in Eqs.~(4), we urgently point to the formal analogy of the first
equation there with Ohm's law, as this is a major premise of our current approach.

Now we are ready to present Fig.~2 that encompasses many of the issues raised
so far. Just the upper funnel in Fig.~1 for water inlet is not yet used here.
At $t = 0$, the hydrated ions begin to fill up the all-important cavity
in Fig.~1 after traversing the blurred zone (see above). Then, as the
concentration in this cavity increases, their degree of hydration descends and
gradually approaches the one of saturated solution. This consumes energy and the
TEG measures the corresponding temperature drop with respect to the reservoir
(note that the warming outside the cavity is not subject to our measurements). The
circles in Fig.~2 represent measured Seebeck voltages normalized to the saturation
concentration $c_0$ at large times, which corresponds to around 0.9~mV or
$\Delta T \approx 0.015$~K across our TEG. After a characteristic time for our
chosen geometry $\tau \approx 155$~s the concentration in the cavity amounts to
$c_0 (1-1/e)$, that is, just a $1/e$~fraction of $c_0$ is still missing.

Based on the just above stated characteristic time $\tau$, we can also plot the
approximate generalized voltage $\Delta c(t) = c_0e^{-t/\tau}$. In view of the
filter paper that acts as a particle inlet barrier, an exponential function with
measured decay constant is much more convenient and trustworthy than direct
calculation of $n({\bf x},t)$ using Eqs.~(4). In sum, we record a spatial
average of concentration-dependent temperature changes (drops due to hydration
decrease) over the TEG area, measured as a momentaneous overall thermovoltage,
which then provides both curves in Fig.~2. After all, at the top right we see
that the vertical error bars (voltage measurement plus inherent process
instability for our setup) are fairly well given by the diameter of the circles,
and at small times the precision is even better.

By virtue of the diffusion dynamics, Fig.~2 depends, e.g., on the dimensions
of the TEG. We think it is worthwhile to look at Fig.~3 that renders things now
on the more natural abscissa of the generalized voltages $\Delta c$. Thus, the
exponential in Fig.~2 degenerates to a straight line with slope unity, as
this is now simply a function of itself. The most important issue is the strongly
unequal distribution of the previously equidistant circles.

\begin{figure}

\vspace{-6.5cm}

\includegraphics[scale=0.65, angle=0]{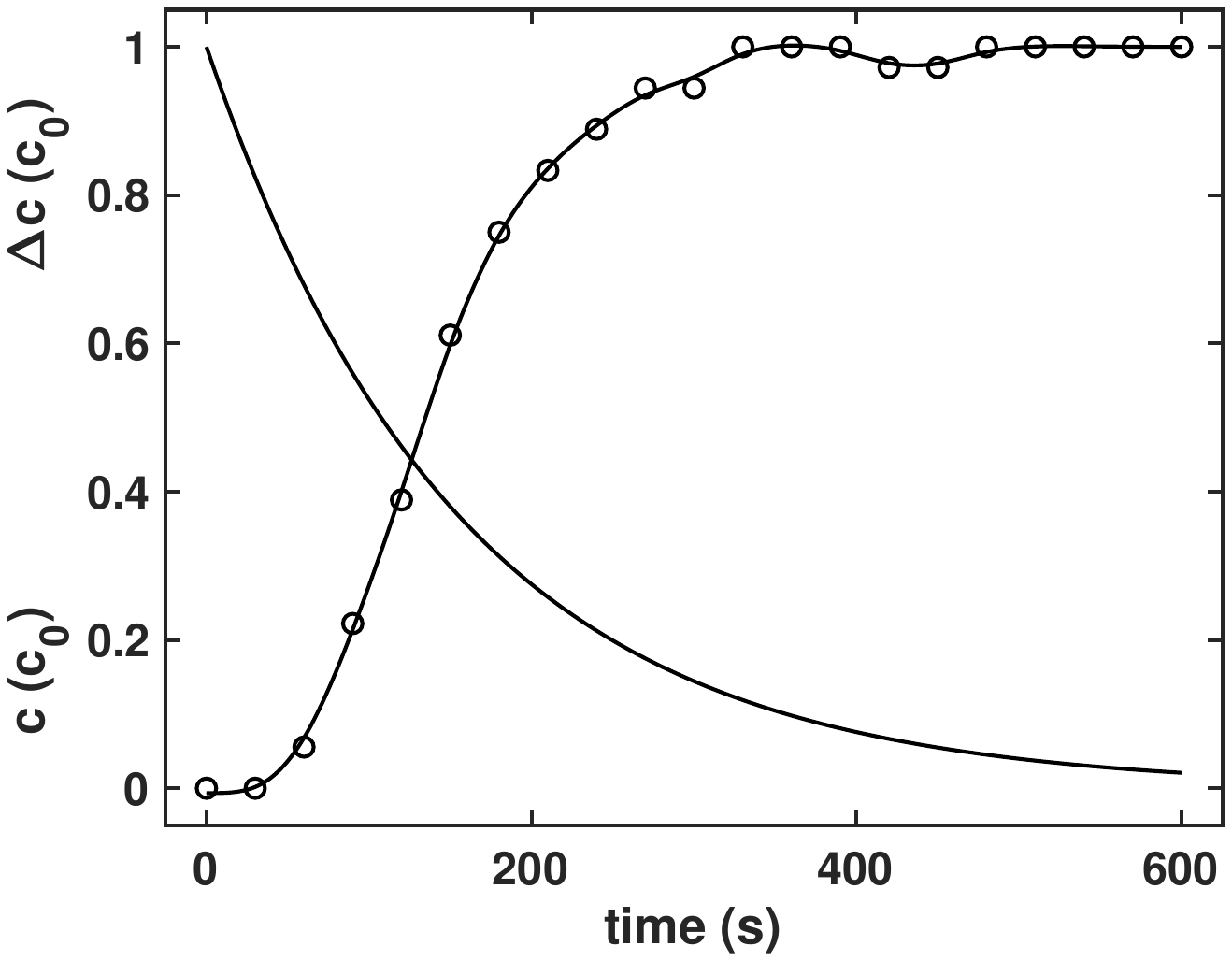}

\vspace{-6.3cm}

\caption{The circles represent measured Seebeck voltages across the TEG normalized
to saturation concentration $c_0$, see text for experimental procedure, efficiency
as well as precision. Higher ordinate values indicate cooling in the cavity above
the TEG through progressing hydration loss of the dissolved Na$^+$ and Cl$^-$ ions.
Also plotted is the exponentially modeled generalized voltage $\Delta c(t)$ with
a characteristic time ($1/e$ decay) of $\approx 155$~s, as obtained from the fitted
curve for the measured TEG voltages converted into concentrations $c(t)$.
\label{fig2}}

\vspace{0.3cm}

\end{figure}

\begin{figure}

\vspace{-6.5cm}

\includegraphics[scale=0.66, angle=0]{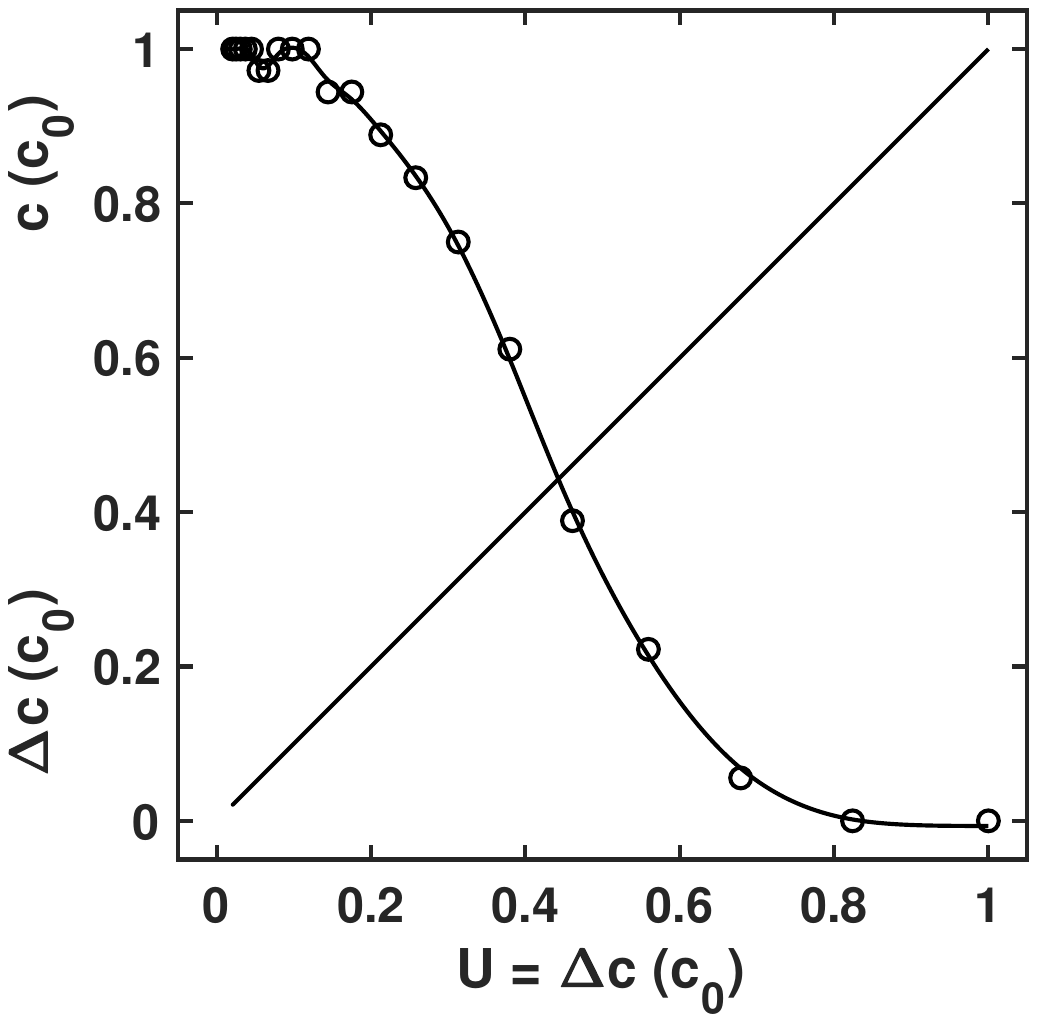}

\vspace{-6.4cm}

\caption{For the sake of transparency, we replot the entire Fig.~2 with the natural
abscissa of the generalized voltages $\Delta c$ that normalizes away much of the
geometrical arbitrariness in the setup. The ``voltages" $\Delta c$ of the ordinate
then are simply a function of themselves. Note on the left that both curves do not
start precisely at zero voltage, as complete concentration equilibration in Fig.~2
is only achieved asymptotically.
\label{fig3}}

%\vspace{0.3cm}

\end{figure}

Only now, in Fig.~4, we come to the analogues of the currents in the Esaki diode,
namely the ordinary diffusive particle flow, and finally the tunnel current that
interests us the most. We conveniently set the volume of the cavity to be unity,
so that we do not need to distinguish particle numbers and concentrations.
Admittedly, our ordinary particle current is somewhat ill-defined in the sense
that the hydrated ions enter the cavity and then do not instantaneously reach its
innermost part, and so, in a simple linear model for the mean flow, we consider
half of the particle entries per unit time to be the actual current. However,
we emphasize that this will have no relevance at all for our subsequent
investigations, and therefore we prefer a straightforward definition of
this certainly less important part of the total current. Then, based on the
assumption that just one water molecule per ion gets released from the
hydration shell (see below), the cooling in Figs.~2 and~3 that is normalized
to saturation concentration $c_0$ measures the number of particles that have
contributed to this thermal effect. The ordinary current is proportional
to our already stated generalized voltage $U(t) = \Delta c(t) = c_0e^{-t/\tau}$,
and the integral of this current over all times for our above plain model amounts
to $c_0/2$, which is half of the saturation totality $c_0$ of particles
that finally have arrived in the cavity (recall that we equate particle
numbers and concentrations, as the volume is thought to be unity). Altogether,
the ordinary (or ohmic in our generalized perception) current as a function of
time reads
\begin{equation}
I_{ohmic}(t) = \frac{c_0}{2\tau} e^{-t/\tau} = \frac{U(t)}{2\tau},
\end{equation}
which is normalized as
\begin{equation}
\int^\infty_0 I_{ohmic}(t) dt = \frac{c_0}{2},
\end{equation}
and in Fig.~4 we merely need the proportionality factor or actually conductance
$1/(2\tau)$ in $I_{ohmic}(U)$ from the second part of Eq.~(5). In the general
case, that is, no arrangement specified, this factor corresponds to the diffusion
constant $D$ in Eqs.~(4). This is intuitive, since a lower characteristic
time $\tau$ increases the diffusion rate in Fig.~2 and also means an enlarged
factor $1/(2\tau)$.

Next we proceed to our major concern in this article, namely the generalized
tunnel current. On this score, we should care for the concentration-dependent
degree of hydration of the ions. What really matters is the increase in hydration
as the ion concentration drops and vice versa, while the actual number of attached
water molecules affects Fig.~4 only quantitatively. It is not that easy to extract
the quantitative extent of hydration from the literature, since the determined
hydration status of an ion significantly depends on experimental as well as
computational methods. We present a choice of references that have particularly
attracted our interest, namely (in chronological order) Refs. \cite{soper,onori,
barnett,rempe,carrillo,korolev,sahle}. These findings may be summerized
very briefly as follows: most authors obtain a primary shell with more
strongly bound water molecules, namely $4 \pm 1$ and $2 \pm 1$ of them around
Na$^+$ and Cl$^-$, respectively, and some authors do not distinguish inner and
outer shells. In addition, an upper limit of involved water molecules is somewhat
arbitrary, as this depends on where the too-weak interaction is truncated.

\begin{figure}

\vspace{-6.5cm}

\includegraphics[scale=0.65, angle=0]{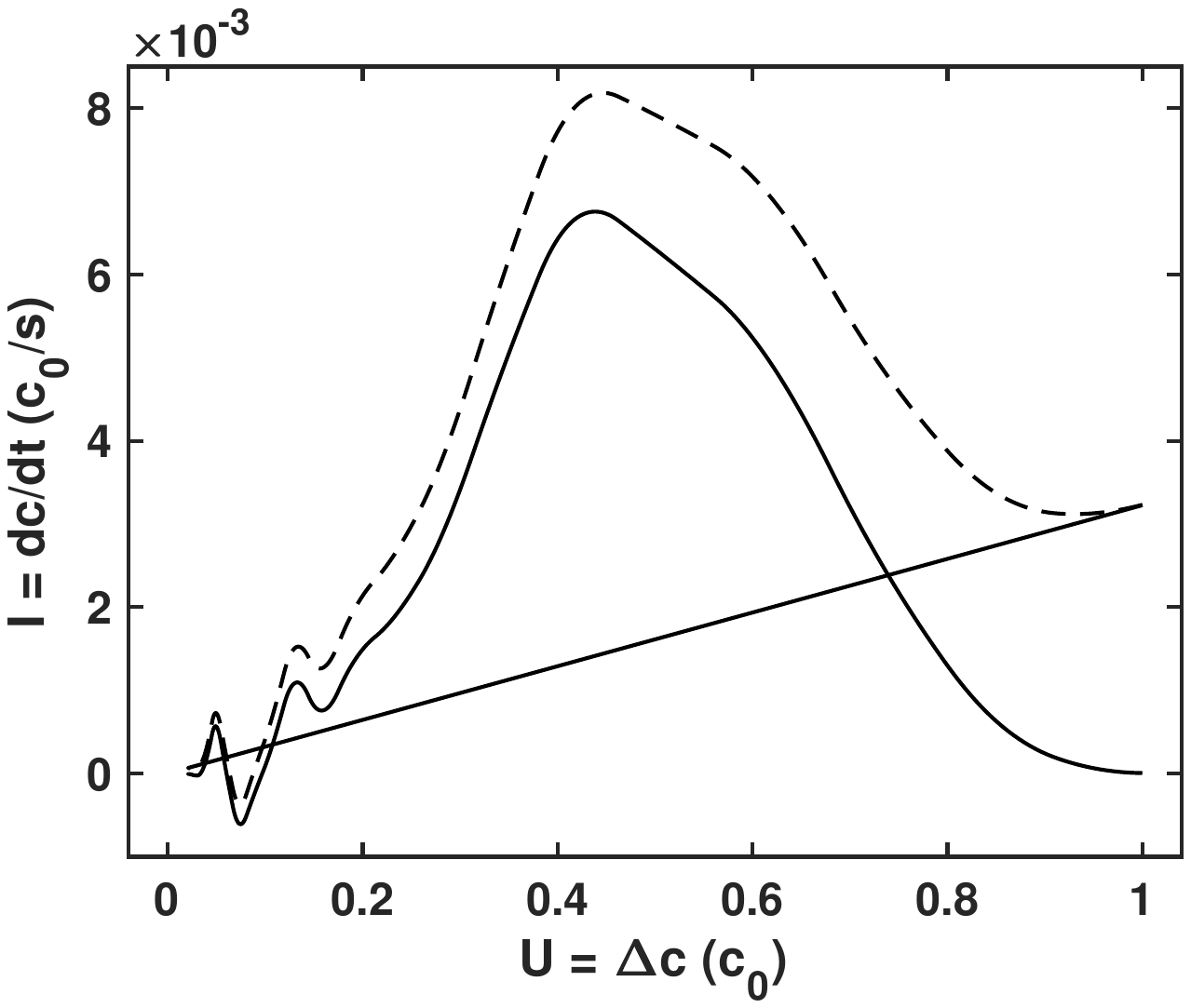}

\vspace{-6.3cm}

\caption{The two full curves are the ordinary or ohmic current (straight line) and
tunnel-only current, respectively, as described in the text. Dashed curve: the sum
of the two said currents, that is, the analogue of the total forward current in an
electronic tunnel diode which, however, has a nonlinear $I(U)$ characteristic for
the ordinary current share. The undulation to the left is due to minor stability
problems visible in Fig.~2 and becomes accentuated by the time derivative.
\label{fig4}}

\vspace{0.3cm}

\end{figure}

As already mentioned, we assume that, averaged over all the ions, only a single
water molecule per ion finally gets released in the cavity. This number (one
molecule per ion) depends on the efficiency of our setup, e.g., how well the
blurred zone outside the cavity (see above) is able to complete the hydration
of ionic particles arriving from the reservoir. More importantly, it strongly
matters where in the hydration shell the molecules come from, and so we actually
should weight them accordingly. Thus it is not really useful to question this
number, as it depends on how it was determined. In this respect, Fig.~4 might then
be subject to quantitative changes.

And so, the generalized tunnel current is just the time derivative of the
momentaneous number of particles accumulated in the cavity (that has unit volume,
see above) and have already lost a water molecule. We also plot the
sum (dashed line in Fig.~4) of ohmic and tunnel current in order to show
the structural analogy to the ``true" electronic tunnel diode. However,
this is not really what we are after. Our main subject of investigation
is the tunnel current only, and by virtue of the measured hydration
heat, this may well be separated from the ohmic contribution that is an indirectly
obtained particle flow. This is in fact a qualitative difference from the original
Esaki diode that exhibits an electric tunnel current without such aforesaid
thermal effects, just as in the conventional current share based on conduction
mechanisms of any semiconductor diode. Note also that more than half of the
abscissa in Fig.~4 is attended with a negative differential resistance $dU/dI$ for
the tunnel current alone curve.

The thermal effects are completely decoupled from the ohmic governed particle
current, since the released and now independent water molecules in the cavity
did not come in as individual particles from the outside, but were an attached
part of a hydrated ion. However, this is of paramount importance, as this
implies a non-dissipative heat flow out of the cavity, giving rise to the
measured cooling inside. Hence, the range of negative differential resistance
may be harnessed to undamp thermal oscillations that do not dissipate
energy from the driving power supply. Clearly, this range should never be
exceeded by the (generalized) voltage or current amplitude in the spirit
of Fig.~4, and so we must permanently harvest the thermal oscillation energy.
We think this is a possible second law of thermodynamics violation, as the said
heat flow out of the cavity spontaneously constitutes a temperature difference,
and the input is only the externally powered particle concentration gradient that
does not involve the thermal energy attended with unbalanced temperatures. This
must then be organized recurrently in order to maintain the oscillations.

We show this relationship between thermal and concentration oscillations in our
Fig.~5, which this time also involves the upper funnel in the Fig.~1 setup. Again,
the circles are measured points, and the fit still features moderate smoothing,
as is easiest seen at the tops of the three structures (maximum cooling in the
cavity). The steep rises indicate the flushing events, that is, we pour 0.2~cm$^3$
of plain water into the upper funnel. The basic level, thus apart from the peaks,
compares to the 0.9~mV of Fig.~2, but the three peak heights indicate lower
temperatures. This is expected, as the liquid in the cavity is somewhat cooler
than the one right above the TEG, where thermal conduction out of the reservoir
has an influence (however, thermal equilibration in Fig.~2 takes place on a much
longer time scale). The cooler saltwater from the bulk region of the cavity then
flows over the TEG and causes the measured ``cold shocks". A larger amount of
flushing water results in even stronger shocks and, also importantly, improves
the salt depletion in the blurred zone (see above) outside the cavity. But, in
a cautious attitude, we stay with an amount somewhat less than the cavity volume.

\begin{figure}

\vspace{-7.0cm}   %{-6.5cm}

\includegraphics[scale=0.65, angle=0]{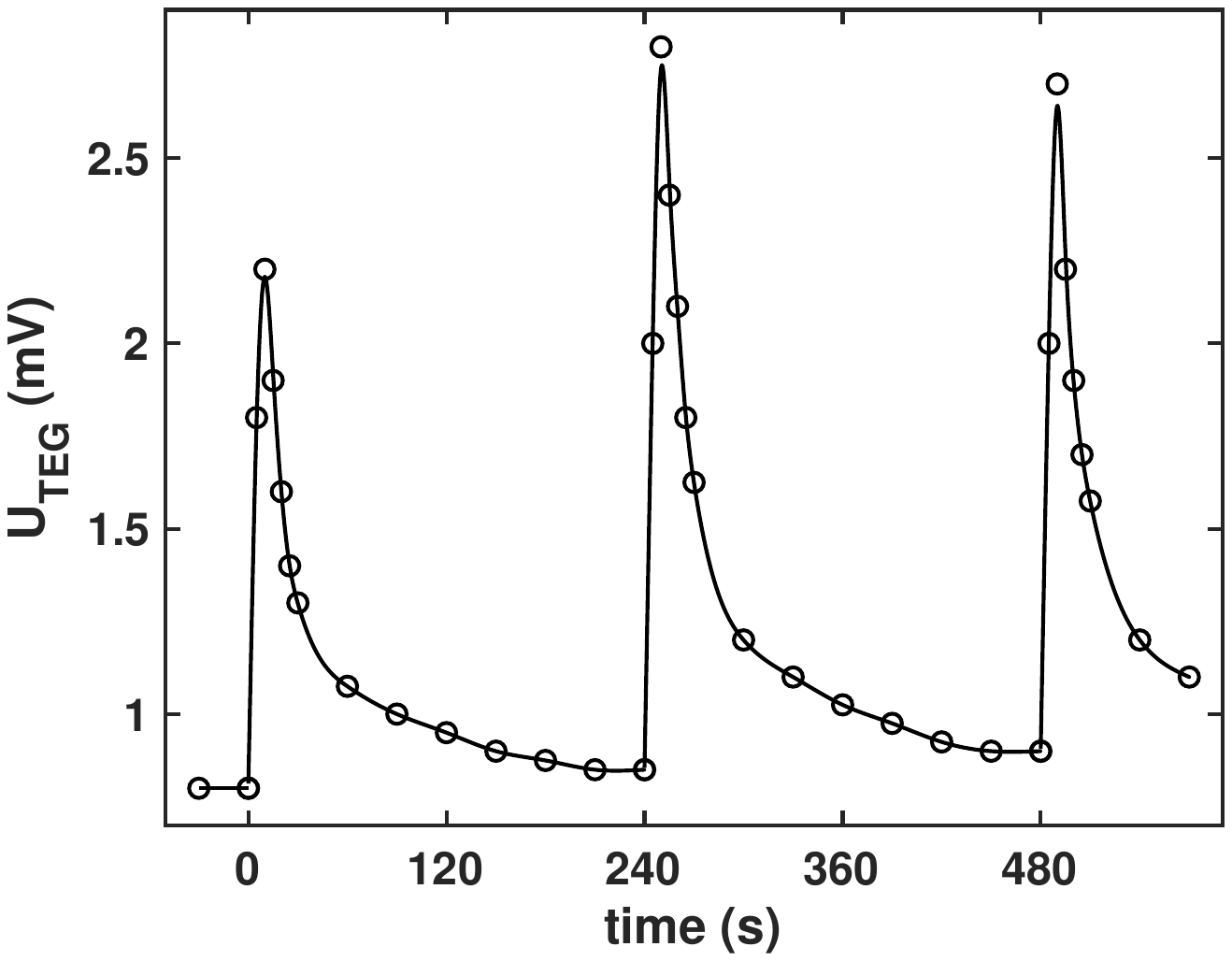}

\vspace{-6.5cm}   %{-6.3cm}

\caption{The three steep ascents are flushing processes at well-taken time lags in
order to recurrently restore the initial situation, see text. This time we directly
record the TEG voltages that measure the cooling with respect to the reservoir,
where every flushing event brings along the cooler liquid from the interior or bulk
area of the cavity under consideration. Again, the circles are measured points and
the fit comprises moderate smoothing, thus slightly understates the peak heights.
\label{fig5}}

\vspace{0.3cm}

\end{figure}

The actual peak heights depend, among other things, on the current conditions in
the upper funnel. Further, even their long-term average is of minor importance,
since this also depends on the geometrical details in the overall setup. Fig.~5 is
just intended to demonstrate how the initial situation with a concentration drop
between reservoir and the cavity interior (extended by the blurred zone) can be
organized in a periodically recurrent manner.

Moreover, it may be helpful to mind that the situation in Fig.~5 withstands
a complete reversal of concentrations, that is, the reservoir now contains plain
water, and the cavity as well as lower funnel are initially filled with saturated
NaCl solution. Most importantly, ``flushing" is also performed with saturated
solution. Then, the liquid inside the cavity gets diluted upon concentration
equilibration and thus warms up. The flushing processes cause the warmer liquid
from the more central area of the cavity to flow over the TEG again, and so we
experience ``heat shocks" of similar magnitude as the previous coolings.

We have already touched on the possibility of exceptions to the second law, and
therein we disregard the issues of a Maxwell's demon and of time reversal. To
our knowledge, there are no such exceptions in statistical mechanics based on
point-like particles, that is, in a somewhat hypothetical world where these
points are nevertheless realistically spaced for fluids or solids (the ideal gas
is a common example). Interestingly, the domain of continuum mechanics (no inner
structure or constituents) admits extensions where these exceptions or violations
can occur, see \cite{ostoja}. In the present article, however, we pursue quite
another course, namely we scrutinize some rather hidden properties of our working
medium, hence of the dissolved hydrated Na$^+$ and Cl$^-$ ions.

Regarding particle size only, we summarized our ideas in \cite{moser}, where we
studied a porous medium with hollow sizes that are reasonably related to the
particle extent. There we experienced small but undoubtedly measurable entropic
consequences. What is most striking in such measurements is the dependence of the
diffusion constant $D$ on the geometrical conditions, as is shown particularly
clearly in \cite{lancon}. Essentially, we dealt with ``thermodynamics" of small
systems, namely the pores or hollows in such media, and these weakly coupled
subsystems then add up to a macroscopic volume. For an interesting perspective on
such small systems, see \cite{sekimoto} and very recently \cite{watanabe}. We
point to linear response theory that provides a convenient way to theoretically
tackle such issues, as the forces of the inner walls on the particles may be
naturally incorporated into a generalized external (scalar) force $f(t)$. Let
$x(t)$ be a (for the sake of simplicity again scalar) unperturbed system's signal.
We state the progressing time-average of its response to said generalized force as
\begin{equation}
\langle x_{resp}(t) \rangle_t = \int^\infty_{-\infty} \Theta(t-\tau)
\chi(t-\tau) f(\tau) d\tau.
\end{equation}
Then, the Fourier transform $\tilde{\chi}(\omega)$ of the linear response function
or susceptibility $\chi(t)$ in Eq.~(7) enters
\begin{equation}
\tilde{P}(\omega) = \frac{2k_B T}{\omega} \rm{Im} \tilde{\chi}(\omega)
\end{equation}
that is the fluctuation-dissipation theorem, see \cite{kubo,blundell}. The
Heaviside function $\Theta(t-\tau)$ in Eq.~(7) assures causality with respect
to the effect of $f(t)$, and $\tilde{P}(\omega)$ in (8) means the Fourier
transform of $x(t)^2$ or power spectral density. For the Brownian particles that
make up diffusion, theorem (8) states that the dissipation caused by $f(t)$, thus
essentially $\rm{Im} \tilde{\chi}(\omega)$, has the same physical origin as the
Brownian fluctuations (characterized by the $\tilde{P}(\omega)$ distribution)
themselves. Further, depletion effects are a fairly known research domain where
particle size may cause entropic phenomena, see, e.g., \cite{karzar}. Clearly, in
our present approach there are no such peculiarities of porous media (apart from
the diffusion barrier between reservoir and cavity, see above), and also depletion
effects are minor. However, we point to the strong overlap with our current work,
as nonzero particle size is really crucial. A theory where the water molecules
for a hydration shell just disappear into point-like particles is, in our view,
somewhat strange.

A final remark is devoted to Jarzynski's equality \cite{jarzynski} that bypasses
the inequality in the second law of thermodynamics; in plain terms, entropy tends
to increase (or remains constant in the Hamiltonian, that is, conservative limit).
The statement reads
\begin{equation}
\langle e^{-W/k_B T} \rangle = e^{-\Delta F/k_B T}
\end{equation}
and relates the Helmholtz free energy difference $\Delta F$ to the work $W$
supplied to a system, regardless whether or not we slowly pursue a reversible path
of quasi-static states starting at equilibrium situation of reservoir temperature
$T$. Here, in $F = U - TS$ with $U$ being the internal energy, the term due to
dissipation $TS$ of entropy $S$ may well grow. The inequality (entropy remains or
rises, or also $\langle W \rangle \ge \Delta F$) is removed, and this actually
supports strict validity of the second law that this way might relate to other
physical statements. Jarzynski's equality goes on to be an active research field,
and exciting generalizations have also emerged, see, e.g., Ref.~\cite{knipschild}.
However, just as in the particle size discussion above, this is not our primary
goal, since we tackle the scope of the second law from a different perspective.
By virtue of hydration, our particles have changed size and mass along their
way, and this is not covered in Eqs.~(8) and (9), neither are the attended thermal
effects that are the major subject of this article. Nevertheless, the particles
are always small enough that equipartition holds true.

\section{Conclusions and outlook}
Overall, we have investigated an analogue of the tunnel diode, with the driving
force now being a concentration gradient of dissolved ionic particles (we used
NaCl in the present work). Clearly, to reproduce the negative differential
resistance is imperative, and we point out its long and steep range in Fig.~4.
Diffusion processes are notoriously slow, but we envision thermal applications
where this is not overly disadvantageous.

The tunnel current analogue consists of water molecules of a hydration shell that
in a sense are ``stowaways", as they later get released in a cavity (upon growth
of concentration therein). But this process, only after a maximum transfer
rate of such water molecules, gradually comes to an end due to concentration
equilibration, which also explains the negative differential resistance.

Here it is of great importance that the thermal hydration effects only concern the
generalized tunnel current, and not the usual (in our analogy ohmic) particle flow
attended with dissipation. This is a beneficial additional feature of our approach
compared to the original Esaki diode where the tunnel current cannot be separated
from the conventional forward current. Definitely this has entropic consequences,
as the resulting heat flow out of the cavity (see above) is opposite to the
inward particle flow driven by a concentration gradient. To our knowledge, the
interdependence between hydration heat and a negative differential resistance has
not yet been investigated, or maybe rather marginally and indirectly.

There is certainly a lot of potential to improve our setup, above all since we
used the energy in a hydration shell only to a small extent. On the theoretical
side, we think statistical mechanics of such over time varying hydrated ions has
not been studied particularly well either, as hydration involves both statistics
and deterministic dynamics. Molecular dynamics simulations may be a promising tool
to anticipate the theory, but there are still many unsettled questions.

\vspace{5mm}
\noindent
{\bf Acknowledgments}

The author thanks P. Robmann and U. Straumann for constructive discussions.

%\begin{acknowledgments}
%\end{acknowledgments}

%\pagebreak
%\vspace{4cm}

%\begin{widetext}
%\end{widetext}

%\clearpage

\end{document}